\documentstyle[12pt]{article}
\begin{document}
\voffset -2.5cm
\textwidth=16cm
\textheight=22cm
\begin{titlepage}
\pagestyle{empty}
\begin{flushright}
\vbox{
{\bf TTP 93-32}\\
{\bf MZ-TH/93-35}\\
{\rm December 1993}\\
\begin{center}
{\large QCD CORRECTIONS TO DECAYS OF POLARISED\\
CHARM AND BOTTOM QUARKS}
\end{center}
\vskip1.0cm
\begin{center}
{\bf A. Czarnecki${}^{a}$, M. Je\.zabek${}^{b,c}$,
J.G. K\"orner${}^{a}$ and J.H. K\"uhn${}^{c}$}
\end{center}
\vskip0.5cm
\noindent
${}^a$ Institut f. Physik, Johannes Gutenberg-Univ.,
D-55099 Mainz, Germany\\
${}^b$ Institute of Nuclear Physics, Kawiory 26a, PL-30055 Cracow,
Poland\\
\hyphenation{Germany}
${}^c$ Institut~f.~Theor.~Teilchenphysik, Univ.~Karlsruhe, D-76128
Karlsruhe, Germany \\
\vskip1.0cm
\begin{center}
Abstract
\end{center}
\hyphenation{analysers}
Distributions of charged
leptons in semileptonic $\Lambda_c$ and $\Lambda_b$
decays can be used as spin analysers for the corresponding
charm and bottom quarks. QCD corrections to the energy spectra
of charged leptons
have been given  in the literature as well as the corrections
to the angular dependence for polarised up-type quarks.
The analogous
formulae are derived also for polarised down-type quarks.
These results are applied to the
decays
of polarised charm and bottom quarks. The corrections
to the asymmetries are found to be small for charm decays.
For bottom decays they exhibit a non-trivial dependence on the
energy of the charged lepton.

\end{titlepage}

Charmed and beautiful $\Lambda$ baryons offer a unique
opportunity to measure the polarisation of heavy quarks
produced in $Z$ boson decays.
It has been proposed long ago that weak hadronic
\cite{bjorken,jrr,koerner1}
and semileptonic \cite{zerwas,koerner2,BKKZ,mele92,wise}
decays of charmed and beautiful $\Lambda$ baryons can be used as
spin analysers for the corresponding heavy quarks.
The polarisation transfer from a heavy quark $Q$ to the
corresponding $\Lambda_Q$ baryon is 100\% \cite{CKPS}, at least in the
limit
$m_Q\rightarrow\infty$.
In view of
the growing sample of $\Lambda_c$ and $\Lambda_b$ baryons
produced at LEP there arises an opportunity to measure the
polarisation of the $c$ and $b$ quarks originating from the
decays of the $Z$ bosons.  The angular distributions of charged
leptons from semileptonic decays of $\Lambda_c$ and $\Lambda_b$
can be used to this end.  According to the Standard Model
the degree of longitudinal polarisation is fairly large,
amounting to
$\langle P_b\rangle = -0.94$ for $b$
and $\langle P_c\rangle = -0.68$ for $c$ quarks~\cite{kz}.
The polarisations depend weakly on the angle. QCD corrections
to the Born results have been calculated recently \cite{KPT}.
In the present article
perturbative QCD corrections are discussed to the distributions
of charged leptons produced in decays of polarised
charm and bottom quarks. In the following discussion
the mass of the charged lepton is neglected,
i.e. the results of the present work can be used for
electrons and muons produced in semileptonic decays
of charm and bottom quarks.
The analogous calculation is in progress
for $\tau$ leptons in the final state.

It is well known~\cite{cm} that the QCD corrections
modify significantly the lifetimes of heavy quarks.
The shapes of the lepton energy distributions, however,
are not strongly affected by QCD
corrections~\cite{ali,ccm,corbo,altar,jk1}.
This is a consequence of the fact that
the  correction function is fairly flat
away from the energy endpoint.
Thus, the correction can be absorbed
into the overall normalisation.
However, in the decay of polarised quarks the polarisation
dependent and independent parts may be affected in different
ways. In such a case the precision of the polarisation measurement
would be spoiled. Therefore, an evaluation of QCD corrections
is necessary.

QCD corrections to the energy spectra of charged leptons
in the decays of $c$ and $b$ quarks
have been calculated in ref.~\cite{ali,ccm} and \cite{ali,corbo},
respectively.
In a later calculation \cite{jk1}
errors were traced in \cite{ccm,corbo,altar}.
The results of \cite{jk1} agree
with the Monte Carlo for charm and bottom \cite{ali}
and the numerical results for top \cite{jk}.
In \cite{jk} the formulae  from~\cite{ali}
were used, and the accuracy achieved was good enough
to observe discrepancies with \cite{ccm} and \cite{corbo}.
After a trivial integration over the angle between the
spin vector of the decaying quark and the direction of
the charged lepton the results described in the present paper
reproduce the results of \cite{jk1} adding a further cross-check
to those already provided there.

The QCD corrections to the joint angular and energy
distributions of charged leptons in polarised top quark decays
have been calculated in~\cite{czajk91}. In the present article
these results are applied and the
angular dependence is evaluated
for charged leptons in charm quark decays.
The corresponding QCD corrections
have been calculated also for the polarised bottom quark.
The calculations follow closely the method outlined
in refs.~\cite{jk1} and \cite{czajk91}.
In particular the notation is the same and the calculations
are performed in the rest frame
of the decaying quark of mass $m_1$.
The angle between its polarisation vector
$\vec s$ and the three-momentum of the charged lepton
is denoted by $\theta$ and $x = 2E_l/m_1$ is
the scaled energy of the charged lepton. The variable
$x$ varies between 0 and $x_{_M}=1-\epsilon^2$,
where $\epsilon=m_2/m_1$ and
$m_2$ denotes the mass of the quark produced in the decay.
$S\equiv|\vec s|=1$ corresponds to fully polarised,
$S=0$ to unpolarised decaying quarks.
The square of the invariant mass of the leptons
in the final state is denoted by $y=\mu^2/m_1^2$.

In Born approximation
the double differential distribution for polarised
charm quark decays
is given by the following formula
\begin{eqnarray}
{{\rm d}\Gamma^{(0)}_c \over {\rm d}x\,{\rm d}\cos\theta  }\,
=\, {G_{_F}^2 m_c^5\over32 \pi^3}\,   {x^2(x_{_M}-x)^2\over 1-x}\,
(1 + S\cos\theta)
\label{cpolborn}
\end{eqnarray}
For the polarised bottom quark one has (see e.g. \cite{tsai79})
\begin{eqnarray}
{{\rm d}\Gamma^{(0)}_b \over {\rm d}x\,{\rm d}\cos\theta  }\,
=\,{G_{_F}^2m_b^5\over32 \pi^3}\, {x^2(x_{_M}-x)^2\over 6(1-x)^2}
\left[ 3-2x+\epsilon^2+{2\epsilon^2\over 1-x}\right.
\nonumber\\
\left.+ S\cos\theta\,\left(1-2x+\epsilon^2
-{2\epsilon^2\over 1-x}\right)\right]
\label{bpolborn}
\end{eqnarray}

The QCD corrections to the above formulae arise from the
exchange of virtual gluons and from real gluon radiation.  In the
calculation of the virtual effects we follow the classic articles
on muon decay~\cite{radcor}.  The contribution of real gluon
radiation is calculated exactly in the same way as in
ref.~\cite{czajk91}.  The final result is free from infrared
divergences and can be expressed as follows:
\begin{eqnarray}
{{\rm d}\Gamma^{(1)}_{c,b} \over {\rm d}x\,{\rm d}\cos\theta
}\,=\, -{G_{_F}^2m_{c,b}^5\over 32 \pi^3}\, {2\alpha_s\over
3\pi}\,\int_0^{y_m} {\rm d}y\,(F_1^{c,b}\, +\, S\cos\theta
J_1^{c,b})
\label{bspec}
\end{eqnarray}
where
$y_m=x(x_{_M}-x)/(1-x)$.  The lengthy expressions for the
coefficient functions $F_1^{c,b}$ can be found in
\cite{jk1}\footnote{The formulae for the neutrino spectrum in
up-type quark decay describe the charged lepton
spectrum for a
down-type quark.}, and even more lengthy ones for $J_1^c$ can be
extracted from \cite{czajk91}.
The coefficient functions $J_1^b$
as well as technical details of the calculation for the polarised
$b$ quark will be given in a separate publication \cite{cj93}.
After numerical integration the correction to the double
differential distribution can be cast into the following form
\begin{eqnarray}
{{\rm d}\Gamma^{(1)}_{c,b} \over {\rm d}x\,{\rm d}
\cos\theta }\,=\,
{{\rm d}\Gamma^{(0)}_{c,b} \over{\rm d}x}\,
\left(-{2\alpha_s\over 3\pi}\right)\,
{1\over2}\left[\,G^{c,b}(x)\,+\,S\cos\theta\,
G^{c,b}_{\rm s}(x)\,\right]
\label{GGs}
\end{eqnarray}

The distributions ${{\rm d}\Gamma^{(0)}_{c,b} / {\rm d}x}$
are well known and follow trivially from
eq.~(\ref{cpolborn}) and eq.  (\ref{bpolborn}).  The functions
$G^{c,b}$ and $G_{\rm s}^{c,b}$ are shown in Figs.~1 and 2 for charm
and bottom respectively.
\begin{figure}
\vskip 7.0cm
\caption{The correction functions a) $G^{c}$ and b) $G_{\rm s}^{c}$
for charm quark:
$\epsilon$=0.3, 0.1 and 0.0 -- solid, dash-dotted and dashed lines.}
\label{Gcharm}
\end{figure}
For charm the numerical values
$\epsilon=0$, $0.1$ and $0.3$ are adopted to bracket the range
of mass values conceivable for Cabbibo allowed and suppressed
decays. For bottom the values $\epsilon=0$ and $0.35$ have been
used corresponding to $b\to u$ and $b\to c$ transitions.
The functions $G^{c,b}$ have been given in \cite{jk1}.
They are plotted in Figs.~1a and 2a for completeness.
In Figs.~1b and 2b the corresponding results are shown for the
functions $G_{\rm s}^{c,b}$.
\begin{figure}
\vskip 7.0cm
\caption{The correction functions a) $G^{b}$ and b) $G_{\rm s}^{b}$
for bottom quark:
$\epsilon$=0.35 (solid) and $\epsilon$=0.0 (dashed).}
\label{Gbottom}
\end{figure}
Both $G^{c,b}$ and $G_{\rm s}^{c,b}$ exhibit
the characteristic endpoint singularity.  This singularity is
closely related to the infrared divergences and consequently
follows the behaviour of the Born matrix element; cf.~the
discussion in \cite{altar,jk1}.  Thus the endpoint singularities
will cancel in the ratio relevant for the correction to the
asymmetry parameter discussed below.
Away from the endpoint the QCD corrections to the angular
dependent and independent parts are typically of the order of 20\%.
However, for charm decays as well as for $b\to c$ transitions
they also cancel to a large extent
in the ratio.
This cancellation, which we believe to be a nontrivial and
important result of the present calculation, has been also
observed in the case of top quark decays \cite{czajk91}.

In analogy to the description of muon decay
we define the asymmetries
$\alpha_{c,b}(x)$ which characterise the angular distributions of
leptons in the semileptonic $c$ and $b$ quark decays:
\begin{eqnarray}
{{\rm d}\Gamma_{c,b} \over {\rm d}x\, {\rm d}\cos\theta  }\,=\,
{{\rm d}\Gamma_{c,b} \over {\rm d}x}\,
\left[\,1 \pm\alpha_{c,b}(x) \cos\theta\,\right] \,/2
\label{five}
\end{eqnarray}
The $\pm$ sign in the definition (\ref{five}) corresponds to
to the weak isospin $I_3 = \pm 1/2$ of the decaying quark\footnote{
The minus sign for $b$ decays follows
the convention for muon decays,
see e.g. ref.~\cite{colphys}. }.
In the Born approximation the asymmetry function for charm
is independent of $x$:
\begin{equation}
\alpha_c^{(0)}(x)=S
\end{equation}
The corresponding asymmetry function for bottom is given by
\begin{equation}
\alpha_b^{(0)}(x)\,=\,-S\,{(1-x)(1-2x+\epsilon^2)
- 2\epsilon^2\over
(1-x)(3-2x+\epsilon^2) + 2\epsilon^2}
\end{equation}

In the following we assume that the decaying quark is
completely polarised, i.e. $S=1$.
The QCD correction to $\alpha_c(x)$ is small. For $\alpha_s=0.3$
and $\epsilon\leq 0.3$ it decreases the Born value
$\alpha_c^{(0)}(x)=1$
by less than 1\%~. The case of $b$ quark is more complicated.
The QCD corrected asymmetry function
$\alpha_{b}(x)$, see solid lines in Fig.3a, and its Born
approximation
$\alpha_b^{(0)}(x)$, dashed lines,
are non-trivial functions of $x$.
The correction to the asymmetry
$\alpha_b(x)\,-\,\alpha_b^{(0)}(x)$,
see Fig.3b, is important in the region where
the Born asymmetry
$\alpha_b^{(0)}(x)$
is small. For $\epsilon=0$ this correction is non-negligible
in a much broader range of $x$.
However, in the most interesting case of $b\to c$
transitions ($\epsilon=0.35$) the effect of the QCD correction
to the asymmetry function
$\alpha_b(x)$
is fairly small.
\begin{figure}
\vskip 6.0cm
\caption{The asymmetry functions for bottom quark, $\alpha_s$=0.2,
$\epsilon$=0 and 0.35~:
a) QCD corrected $\alpha_b(x)$ -- solid -- and Born
$\alpha_b^{(0)}(x)$ -- dashed
lines;
b) $\alpha_b(x)\,-\,\alpha_b^{(0)}(x)$ for the transitions
$b\to c$ ($\epsilon=0.35$)
-- solid -- and $b\to u$ ($\epsilon=0.0$) -- dashed line.}
\end{figure}

The degree of polarisation of charmed quarks from $Z$ decays
is smaller than for bottom quarks. On the other hand the
original polarisation of $c$ quarks is reflected in the
experimentally accessible angular distributions of
charged leptons practically
without any loss. This is not the case for bottom quarks.
The angular asymmetry for charged leptons from $b$ decays
is smaller than 1 and even changes its sign as a function
of the charged lepton energy.
Hence both charm and bottom modes are worth pursuing.
In contrast to the charged leptons the double differential
angular and energy distributions of neutrinos
reflect the original polarisation of bottom quarks
without any loss of information.
Thus, the use of neutrinos as spin analysers
may be worth further experimental considerations.

\vskip0.5cm
\par\noindent
{\em Acknowledgements.}
Work partly supported by KBN under grant 203809101,
by BMFT under contracts 056KA93P and 06MZ730,
by DFG and by
Natural Sciences and Engineering Research
Council of Canada under grant OGP3167.

\end{document}